\documentstyle[amssymb,amsmath,aps,preprint,eqsecnum,tighten]{revtex}

\begin{document}
\draft
\title{Null Geodesics in Five-Dimensional Manifolds}
\author{Sanjeev S. Seahra and Paul S. Wesson}
\address{Department of Physics, University of Waterloo,
    Waterloo, ON, Canada N2L 3G1 \\ \tt{ssseahra@uwaterloo.ca}}
\date{\today}
\maketitle

\setlength\arraycolsep{2pt}
\newcommand*{\di}{\partial}

\begin{abstract}

We analyze a class of 5D non-compact warped-product spaces
characterized by metrics that depend on the extra coordinate via a
conformal factor.  Our model is closely related to the so-called
canonical coordinate gauge of Mashhoon et al.\ We confirm that if
the 5D manifold in our model is Ricci-flat, then there is an
induced cosmological constant in the 4D sub-manifold. We derive
the general form of the 5D Killing vectors and relate them to the
4D Killing vectors of the embedded spacetime.  We then study the
5D null geodesic paths and show that the 4D part of the motion can
be timelike --- that is, massless particles in 5D can be massive
in 4D. We find that if the null trajectories are affinely
parameterized in 5D, then the particle is subject to an anomalous
acceleration or fifth force. However, this force may be removed by
reparameterization, which brings the correct definition of the
proper time into question. Physical properties of the geodesics
--- such as rest mass variations induced by a variable
cosmological ``constant'', constants of the motion and 5D
time-dilation effects --- are discussed and are shown to be open
to experimental or observational investigation.

\vspace{0.1in}


\end{abstract}

\section{Introduction}

The extension of 4D spacetime to higher dimensions is now
commonplace, as in Kaluza-Klein theory (5D), superstrings (10D)
and supergravity (11D).  There is currently a large amount of
interest in brane-world theories with non-compact extra dimensions
serving as a possible route to reconciling the formalisms used to
describe particle and gravitational physics \cite{Ran99,Shi99}.
Recent papers have presented and analyzed new exact solutions of
the 5D vacuum field equations in the context of 4D wormholes
\cite{Dzh99} and Friedmann-Robertson-Walker cosmologies
\cite{Wes00}. The motion of test particles in 5D has been much
studied \cite{Lei73,Kov84,Geg84,Dav86,Fer89,Mas94,Wes95,Mas98%
,Wes99a,Wes99b,Liu00}. The five dimensional geodesic equation can
be reduced to 4D equations of motion and an equation governing the
motion in the extra dimension.  As in 4D general relativity, the
precise form of these formulae depends on whether we are
considering timelike $d{\mathcal{S}}^2 > 0$, null
$d{\mathcal{S}}^2 = 0$ or spacelike $d{\mathcal{S}}^2 < 0$
trajectories, where $d{\mathcal{S}}^2 = g_{AB} dx^A dx^B$ is the
5D arclength\footnote{In this paper, we label 5D coordinates $x^A$
by uppercase Latin indices that run 0
--- 4 with $x^4 = y$.  Lowercase Greek letters run over spacetime
indices 0 --- 3.  We employ units where $G = c = 1$.  The
signature of the 5D metric is $(+----)$ while the signature of the
4D metric is $(+---)$.}. The possibility that the dynamics of
particles in Kaluza-Klein theories could involve spacelike 5D
paths was raised by Davidson and Owen \cite{Dav86}, the argument
being that the 4D part of the trajectory could be a casual curve
$g_{\alpha\beta} dx^\alpha dx^\beta > 0$ even if the higher
dimensional trajectory is acausal.  A common feature of the
derived 4D equations of motion is that they do not appear to be
spacetime geodesics \cite{Wes99b}.  That is, there is in general
an anomalous acceleration in 4D due to the fifth dimension, or
equivalently a fifth force.  This has so far not been observed,
either in local dynamics or cosmology \cite{Wil92}.  Recently, Liu
and Mashoon have interpreted this extra force as being related to
variations in the rest masses of test particles traveling on 5D
timelike, null and spacelike geodesics \cite{Liu00}.

In this paper, we examine the nature of this anomalous
acceleration in detail using a 5D model that is conformally
related to the manifold first introduced by Kaluza \cite{Kal21}.
In section \ref{sec:Unique}, we demonstrate that if the 5D vacuum
field equations are enforced, the 4D part of the metric satisfies
Einstein's equations in the absence of ordinary matter with a
cosmological constant $\Lambda > 0$.  In section
\ref{sec:Killing}, we derive the general form of Killing vectors
of the 5D manifold $\xi^A$ and relate them to Killing vectors of
the 4D $y = \mathrm{constant}$ sub-manifold $\eta^\alpha$.  Higher
dimensional particle dynamics is studied in section
\ref{sec:Geodesic} using the assumption that all trajectories are
5D null geodesics, a choice motivated by the special relativistic
relation $E^2 = p^2 + m^2$. It is then shown that particles can
travel along timelike 4D paths even if $d{\mathcal{S}}^2 = 0$.  We
find that if the parameterization of the 5D null curves is affine,
the 4D part of the trajectory is subject to an acceleration
parallel to the 4-velocity.  However, this anomalous force can be
removed by a parameter transformation, which introduces
ambiguities in defining the 4D proper time. This has important
consequences for the determination of variations in rest mass,
which are discussed in section \ref{sec:Mass}.  In the 5D
Ricci-flat case, we show how rest-mass variations can arise from
an exchange of energy between the particle and the vacuum.  The
topic of section \ref{sec:Conserved} is quantities conserved along
the geodesics, while section \ref{sec:Time} discusses
time-dilation effects associated with different parameterizations
and potential experiments to determine the ``true'' proper time.

\section{The 5D Metric and the Uniqueness of Canonical
Coordinates}\label{sec:Unique}

We will study the geodesic motion of particles in a 5D manifold
using a particularly useful coordinate gauge.  Our choice of
coordinates is based on the 5D canonical metric introduced by
Mashhoon et al.\ \cite{Mas94,Mas98,Wes99a,Wes99b}.  The line
element in canonical coordinates is given by
\begin{equation}\label{canonical:metric}
    d{\mathcal{S}}^2 = \frac{y^2}{L^2} g_{\alpha\beta}(x^{\mu},y)
    dx^{\alpha} dx^{\beta} - dy^2.
\end{equation}
Here, $L$ is a constant introduced to give ${d\mathcal{S}}^2$ the
correct units.  The metric (\ref{canonical:metric}) is general in
the sense that the line element in any 5D manifold may be
expressed in the canonical form via appropriate coordinate
transformations. This choice of gauge results in great algebraic
simplification of the vacuum 5D field equations, which identify
the constant $L$ with an induced 4D cosmological constant via
$\Lambda = 3 / L^2$.

The manifold that we examine in this paper is represented by the
5D line element
\begin{equation}\label{restricted:metric}
    d{\mathcal{S}}^2 = \Phi^2(y) g_{\alpha\beta}(x^{\mu}) dx^{\alpha}
    dx^{\beta} - dy^2.
\end{equation}
This is an example of a so-called ``warped-product space'' which
has received a fair amount of recent attention in the literature
\cite{Ran99,Shi99}.  Here, $\Phi(y)$ is an unspecified function of
$y$, which we call the \emph{conformal prefactor}. We will use the
notation
\begin{equation}
    \Omega(y) \equiv \Phi^2(y)
\end{equation}
where convenient (both notations are common in the literature).
Our 5D model is obviously similar to (\ref{canonical:metric}), but
there are two notable exceptions: we do not restrict $\Phi(y) =
y/L$, and $g_{\alpha\beta}$ is assumed to depend on spacetime
variables $x^\mu$ only.  This metric (\ref{restricted:metric}) is
not general and in fact refers to a set of 5D manifolds with a
certain type of symmetry.  We can elucidate this symmetry by
performing a conformal transformation
\begin{equation}
    g_{AB} \rightarrow \Phi^{-2}(y)g_{AB},
\end{equation}
followed by the coordinate transformation
\begin{equation}
    Y = \int^y \Phi^{-1}(u) \, du.
\end{equation}
The line element $d\hat{\mathcal{S}}$ in the conformal manifold is
then given by
\begin{equation}\label{conformal:metric}
    d\hat{\mathcal{S}}^2 = g_{\alpha\beta}(x^{\mu}) dx^{\alpha}
    dx^{\beta} - dY^2.
\end{equation}
This is the classic form of the metric of a 5D manifold in the
absence of electromagnetic potentials $A^\alpha$ \cite{Kal21}.
Thus the 5D $y$-dependent spaces (\ref{restricted:metric}) are
related to ordinary 4D spaces (\ref{conformal:metric}) via a
simple conformal transformation.

These comments imply that all the information about the conformal
5D manifold is embedded in $g_{\alpha\beta}$.  It is for this
reason that we call $g_{\alpha\beta}$ the 4D \emph{conformal
metric}. Now, the \emph{induced metric} on $y=\mathrm{constant}$
4D hypersurfaces $\Sigma_y$ is
\begin{equation}
h_{\alpha\beta} = \Phi^2(y) g_{\alpha\beta}.
\end{equation}
Because the difference between the two 4D metrics is a
$y$-dependent prefactor, both $g_{\alpha\beta}$ and
$h_{\alpha\beta}$ transform as 4-tensors on $\Sigma_y$ and both
satisfy completeness relations.  We will use $g_{\alpha\beta}$ to
raise and lower indices on 4D objects (for example, the projection
of particle velocities onto $\Sigma_y$). Because all the
$y$-dependence of the induced metric is concentrated in the
conformal prefactor, the 4D Christoffel symbols and derived
curvature quantities defined for each of the 4D metrics are
equivalent and independent of $y$.  For all intents and purposes,
$g_{\alpha\beta}$ is the fundamental quantity on $\Sigma_y$.

In most of this paper we will not assume any particular form of
the 5D field equations.  However, it is useful to make contact
with previous work by assuming, like other authors
\cite{Mas94,Mas98,Wes99a,Wes99b}, that the 5D vacuum field
equations are
\begin{equation}\label{field:equations}
    R_{AB} = 0, \quad A,B=0,1,2,3,4.
\end{equation}
We will now prove that given a metric of the form
(\ref{restricted:metric}) and the field equations
(\ref{field:equations}), then the conformal prefactor $\Phi(y)$ is
determined up to an integration constant and a linear translation
on $y$.

It is well known that the 15 equations (\ref{field:equations}) can
be broken down into a set of 10 Einstein equations, a set of 4
Maxwell equations and a wave equation \cite{Wes99b}.  For the
metric (\ref{restricted:metric}), there are no electromagnetic
potentials ($g_{4\alpha} = 0$) and the scalar field is a constant
($g_{44} = - 1$).  It is then straightforward to extract the 4D
part of (\ref{field:equations}), which yields
\begin{equation}\label{field:1}
    R_{\alpha\beta} = -\tfrac{1}{2} \left( {\Omega}'' +
    \Omega^{-1}{{\Omega}'}^2 \right) g_{\alpha\beta},
\end{equation}
where a prime denotes differentiation with respect to $y$.  This
is from the $A,B = 0,1,2,3$ components of (\ref{field:1}). Here,
$R_{\alpha\beta} = 0$ is the 4D Ricci tensor defined with respect
to either the induced metric $h_{\alpha\beta}(x^\mu,y)$ or the
conformal metric $g_{\alpha\beta}(x^{\mu})$. The $R_{4\alpha} = 0$
parts of (\ref{field:1}) are automatically satisfied because
$g_{4\alpha} = 0$.  The 44-component of (\ref{field:equations})
yields
\begin{equation}\label{field:2}
    {\Omega}'' - \tfrac{1}{2} \Omega^{-1} {{\Omega}'}^2 = 0.
\end{equation}
If we now contract the 4D relation (\ref{field:1}), we obtain the
4D Ricci scalar as
\begin{equation}\label{field:3}
    R = -2 \left( {\Omega}'' + \Omega^{-1}{\Omega'}^2 \right).
\end{equation}
However, the left-hand side is a function of spacetime variables
$x^{\mu}$ while the right-hand side is a function of $y$ only.
Hence, both sides must be equal a constant.  We choose
\begin{eqnarray}\label{field:4}
    R & = & -4\Lambda, \\ \label{field:5}
    \Lambda & = & \tfrac{1}{2} \left( {\Omega}'' + \Omega^{-1}
    {\Omega'}^2 \right).
\end{eqnarray}
Then (\ref{field:1}) gives for the 4D Ricci and Einstein tensors
\begin{eqnarray}\label{Ricci}
    R_{\alpha\beta} & = &  -\Lambda g_{\alpha\beta} = -\Lambda
    \Omega^{-1}(y) h_{\alpha\beta}, \\
    \label{Einstein}
    G_{\alpha\beta} & = & + \Lambda g_{\alpha\beta} = +\Lambda
    \Omega^{-1}(y) h_{\alpha\beta}.
\end{eqnarray}
For observers restricted to $\Sigma_y$ hypersurfaces, these are
the conventional equations of general relativity in the absence of
ordinary matter, but with a finite cosmological constant.
[Equivalently, they describe a vacuum state with a pressure and
density that obeys $p = -\rho$ as in the de Sitter model.]  We
will discuss the experiences of freely-falling observers below in
section \ref{sec:Mass}. Eliminating the first-derivative terms in
(\ref{field:2}) and (\ref{field:5}) yields
\begin{equation}\label{field:6}
    {\Omega}'' = \tfrac{2}{3}\Lambda \quad \Rightarrow \quad
    \Omega(y) = \tfrac{1}{3} \Lambda (y - y_*)^2 + k,
\end{equation}
where $y_*$ and $k$ are arbitrary constants.  Substitution of
(\ref{field:6}) into either (\ref{field:2}) or (\ref{field:5})
demands that $k=0$ for consistency.  We hence obtain the solution
\begin{equation}\label{omega:solution}
    \Omega(y) = \tfrac{1}{3} \Lambda (y - y_*)^2,
\end{equation}
which is unique up to a fiducial value of $x^4 = y$, namely $y_*$.
This means that the 4D conformal prefactor in
(\ref{restricted:metric}) is fixed by the field equations
(\ref{field:equations}).  The absorbable constant $y_*$
notwithstanding, (\ref{omega:solution}) defines what are called
canonical coordinates in the literature
\cite{Mas94,Mas98,Wes99a,Wes99b}.  Also, note that we need to
restrict $\Omega(y) > 0$ to ensure that the 5D metric
(\ref{restricted:metric}) is well-behaved, which means that
$\Lambda > 0$. Hence, the 4D sub-manifold represents de Sitter,
not anti-de Sitter, spacetimes.

So, we have shown that in the case where the 5D manifold is
Ricci-flat there is a unique solution for the conformal prefactor
$\Phi(y)$, which corresponds to the usual 5D canonical metric
(\ref{canonical:metric}).  This solution induces a stress-energy
tensor on $y = \mathrm{constant}$ hypersurfaces consistent with 4D
general relativity in the presence of a non-zero cosmological
constant and in the absence of ordinary matter.

\section{Killing Vectors}\label{sec:Killing}

In this section, we will derive the form of the Killing vectors of
the 5D warped-product space described by the line element
(\ref{restricted:metric}). We write 5D Killing vectors as
\begin{eqnarray}
    \xi^A & = & (\Omega^{-1}\xi^{\alpha},-\xi_4) \\
    \xi_A & = & (\xi_{\alpha},\xi_4),
\end{eqnarray}
where $\xi_\alpha = g_{\alpha\beta} \xi^{\beta}$.  We will need
the Christoffel symbols of the 5D manifold, which we denote by
$\hat{\Gamma}^{A}_{BC}$.  They are:
\begin{equation}\label{Christ}
\begin{array}{lcl}
    \hat{\Gamma}^{\alpha}_{\beta\gamma} & = &
    \Gamma^{\alpha}_{\beta\gamma},\\
    \hat{\Gamma}^{\alpha}_{4\beta} & = &
    \frac{1}{2}\Omega^{-1}\Omega'\delta^{\alpha}_{\beta},\\
    \hat{\Gamma}^{4}_{\alpha\beta} & = & \frac{1}{2} \Omega'
    g_{\alpha\beta}, \\
    \hat{\Gamma}^{\alpha}_{44} & = & \hat{\Gamma}^{4}_{4\alpha} =
    \hat{\Gamma}^{4}_{44} = 0
\end{array}
\end{equation}
with
\begin{equation}
    \Gamma^{\alpha}_{\beta\gamma} =
    \tfrac{1}{2}g^{\alpha\sigma}(g_{\beta\sigma,\gamma} +
    g_{\gamma\sigma,\beta} - g_{\beta\gamma,\sigma}),
\end{equation}
where a comma denotes partial differentiation.

The 5D Killing equation is
\begin{equation}
    0 = \hat{\nabla}_A \xi_B + \hat{\nabla}_B \xi_A,
\end{equation}
where $\hat{\nabla}_A$ is the 5D covariant differential operator.
This equation can be split up into three sets of equations in a
manner analogous to the splitting of $R_{AB} = 0$:
\begin{eqnarray}\label{first}
    0 & = & \hat{\nabla}_\alpha \xi_\beta + \hat{\nabla}_\beta
    \xi_\alpha, \\ \label{second}
    0 & = & \hat{\nabla}_\alpha \xi_4 + \hat{\nabla}_4
    \xi_\alpha, \\ \label{third}
    0 & = & \hat{\nabla}_4 \xi_4.
\end{eqnarray}
From the third equation (\ref{third}), we find
\begin{equation}
    \di_4 \xi_4 = 0 \quad \Rightarrow \quad \xi_4 = \Psi(x^{\mu}),
\end{equation}
where $\di_4 = \di / \di y$ and $\Psi(x^\mu)$ is a 4D scalar
function independent of $y$.  Using this fact and the Christoffel
symbols (\ref{Christ}), equation (\ref{second}) becomes
\begin{equation}\label{3.5}
    \di_\alpha \Psi = -\Omega \, \di_4 (\Omega^{-1}\xi_\alpha),
\end{equation}
where $\di_\alpha = \di / \di x^\alpha$.  We can apply the 4D
covariant derivative $\nabla_\beta$ to this result and note that
$\nabla_\beta\di_\alpha \Psi = \nabla_\alpha\di_\beta \Psi$ to get
\begin{equation}\label{fourth}
    0 = \di_4 \left[ \Omega^{-1} \left( \nabla_\alpha \xi_\beta -
    \nabla_\beta \xi_\alpha \right) \right].
\end{equation}
Now, we can expand and rewrite equation (\ref{first}) to give
\begin{equation}\label{4.5}
    {\pounds}_\xi g_{\alpha\beta} = \Omega ' g_{\alpha\beta} \Psi,
\end{equation}
where
\begin{equation}\label{4.75}
    {\pounds}_\xi g_{\alpha\beta} =\nabla_\alpha \xi_\beta +
    \nabla_\beta \xi_\alpha.
\end{equation}
We will now assume that $\Psi \ne 0$ and show that this leads to a
contradiction unless $\Omega(y)$ has a specific form.  If we take
equation (\ref{4.5}), divide by $\Omega$, differentiate with
respect to $y$ and then contract over the spacetime indices, we
obtain
\begin{equation}\label{fifth}
    \Psi^{-1} \square \Psi = -2\Omega \,
    \di_4(\Omega^{-1}\Omega'),
\end{equation}
where $\square \equiv \nabla^{\alpha}\nabla_{\alpha}$ and we have
made use of (\ref{3.5}).  The left-hand side of (\ref{fifth}) is a
function of $x^\alpha$ only, while the right-hand side is a
function of $y$ only.  By separation of variables, we obtain
\begin{eqnarray}\label{5.5}
    0 & = & (\square + 2k_1) \Psi, \\
    \label{sixth} \frac{k_1}{\Omega} & = & \frac{d^2}{dy^2}\ln\Omega,
\end{eqnarray}
where $k_1$ is a constant.  The second of these formulae
represents a second-order ODE that must be satisfied by the
conformal prefactor in order to find a solution to Killing's
equation with $\Psi \ne 0$.  We can integrate (\ref{sixth}) once
to obtain
\begin{equation}\label{5.75}
    \frac{d}{dy}\ln\Omega = k_1f(y) + k_2,
\end{equation}
where $k_2$ is a constant and
\begin{equation}
    f(y) = \int^y \Omega^{-1}(u)\,du.
\end{equation}
We can also integrate (\ref{3.5}) with respect to $y$ by
introducing an arbitrary dual vector field $\eta_\alpha(x^\mu)$
that is independent of the fifth coordinate. This gives
\begin{equation}
    \xi_\alpha(x^\mu,y) = \Omega(y) \eta_\alpha(x^\mu) - f(y) \Omega(y)
    \di_\alpha \Psi(x^\mu).
\end{equation}
Putting this into (\ref{4.5}) yields
\begin{equation}
    {\pounds}_\eta g_{\alpha\beta} = 2f(y)
    \nabla_\alpha \di_\beta \Psi + g_{\alpha\beta}
    \Psi \di_4 \ln\Omega.
\end{equation}
Contracting and making use of (\ref{5.5}) gives
\begin{equation}
    \Psi^{-1}\nabla^\alpha \eta_\alpha = 2(\di_4\ln\Omega - k_1).
\end{equation}
By separation of variables, we require
\begin{equation}
    k_3 = \frac{d}{dy} \ln\Omega,
\end{equation}
where $k_3$ is a constant.  Solving this equation gives
\begin{equation}\label{special:omega}
    \Omega(y) = \Omega_0 \exp(k_3y).
\end{equation}
So, unless the conformal prefactor is given by the above equation,
it is impossible to solve Killing's equation with $\Psi \ne 0$.
Therefore, we must set $\Psi = 0$ for $\Omega(y) \ne \Omega_0
\exp(k_3y)$.

Setting $\Psi = 0$ in (\ref{3.5}) and integrating with respect to
$y$ yields
\begin{equation}
    \xi_\alpha = \Omega(y) \eta_\alpha(x^\mu).
\end{equation}
Putting this into (\ref{4.5}) with $\Psi=0$ gives
\begin{equation}
    \nabla_\alpha \eta_\beta + \nabla_\beta \eta_\alpha = 0.
\end{equation}
Hence, Killing vectors of our 5D manifold are given by
\begin{equation}
    \xi_A = \big(\Omega(y)\eta_{\alpha}(x^\mu),0\big), \quad
    {\pounds}_\eta g_{\alpha\beta} = 0,
\end{equation}
provided that $\Omega(y) \ne \Omega_0 \exp(k_3y)$.  We remark that
the 5D Killing vectors are simply related to the 4D Killing
vectors $\eta_\alpha$ of the conformal metric $g_{\alpha\beta}$.
This fact will be examined more closely below in section
\ref{sec:Conserved}.

\section{The Trajectory of 5D Null Particles}\label{sec:Geodesic}

The affinely-parameterized geodesics of the above manifold
(\ref{restricted:metric}) can be derived from the variation of the
Lagrangian
\begin{equation}\label{Lagrangian}
    {\mathcal{L}} = \tfrac{1}{2} \left( d{\mathcal{S}} / d\lambda
    \right)^2 = \tfrac{1}{2} \left[ \Phi^2 g_{\alpha\beta}
    k^{\alpha} k^{\beta} - \dot{y}^2 \right]
\end{equation}
provided we choose $k^{A} k_{A} = \mathrm{constant}$. [If we work
in a parameterization where the norm of the 5-velocity is
variable, we need to extremize $\int d{\mathcal{S}}$ instead of
$\int (d{\mathcal{S}}/d\lambda)^2 d\lambda$.] Here, $k^{A} \equiv
dx^A/d\lambda$, $k^{\alpha} \equiv dx^{\alpha}/d\lambda$, $\dot{y}
\equiv dy/d\lambda$ and $\lambda$ is an \emph{affine} parameter.
The momenta are
\begin{equation}
    \begin{array}{rclcl}
    p_{\alpha} & = & \partial{\mathcal{L}} / \partial k^{\alpha}
    & = & \Phi^{2} g_{\alpha\beta} k^{\beta},
    \\ p_4 & = &
    \partial{\mathcal{L}} / \partial \dot{y} & = & -\dot{y}.
    \end{array}
\end{equation}
To get the equations of motion, we can use the Euler-Lagrange
equations:
\begin{equation}\label{Euler:Lagrange}
    \frac{d}{d\lambda} \left( \frac{\partial{\mathcal{L}}}{\partial
    k^A} \right) - \frac{\partial{\mathcal{L}}}{\partial
    x^A} = 0.
\end{equation}
After some algebra, the 4D part of these can be written as
\begin{equation}\label{4D:geodesic:1}
    {}^{(\lambda)}a^\alpha = -2\Phi^{-1}\Phi'\dot{y}k^{\alpha}.
\end{equation}
Here and henceforth, we use the notation ${}^{(z)}a^\alpha$ to
denote the 4D acceleration in the $z$ parameterization:
\begin{equation}
    {}^{(z)}a^\alpha \equiv \frac{d^2 x^\alpha}{dz^2} +
    \Gamma^\alpha_{\beta\gamma}
    \frac{dx^\beta}{dz}\frac{dx^\gamma}{dz}.
\end{equation}
In equation (\ref{4D:geodesic:1}), $z = \lambda$.  Equation
(\ref{4D:geodesic:1}) shows that for an affine parameter $\lambda$
along the path, there is a velocity-dependent fifth force.  The
fifth part of (\ref{Euler:Lagrange}) gives
\begin{equation}\label{Fifth:part}
    \ddot{y} = - \Phi \Phi' k_{\gamma} k^{\gamma},
\end{equation}
which shows that, in general, the particle accelerates in the
fifth dimension.

To continue, we need to choose the type of 5D geodesic we are
dealing with.  In 4D relativity, the relation $p^\alpha p_\alpha =
m^2$ implies that $u^\alpha u_\alpha = 1$ for massive particles.
In 5D, a natural extension of the 4D energy-momentum relation is
$p^A p_A = 0$ with the fifth component of the momentum being
interpreted as the particle mass $p_4 \sim m$.  This implies that
5D trajectories are null, which is the hypothesis that we will
work with in the rest of this paper.  Therefore, let us put
$d{\mathcal{S}}^2 = 0$ or $k_A k^A = 0$ for null paths. With the
Lagrangian chosen as (\ref{Lagrangian}), the equations
(\ref{4D:geodesic:1}) and (\ref{Fifth:part}) are still well
defined. In this case, the metric (\ref{restricted:metric}) gives
\begin{equation}\label{Null:Condition}
    \dot{y}^2 = \Phi^2 k_{\gamma} k^{\gamma}.
\end{equation}
We can use (\ref{Null:Condition}) to substitute for $k_{\gamma}
k^{\gamma}$ ($\ne 1$) in (\ref{Fifth:part}).  This gives
$\ddot{y}/\dot{y}^2 = -\Phi^{-1} d\Phi / dy$, which is solved by
\begin{equation}\label{ydot:solution}
    \dot{y} = K\Phi^{-1}(y).
\end{equation}
Here $K$ is a constant.  We can integrate this result noting that
$K d\lambda = \Phi(y) dy$, so in terms of two other constants
$\lambda_0$ and $y_0 = y(\lambda_0)$, we have
\begin{equation}\label{y:solution}
    K(\lambda - \lambda_0) = \int_{y_0}^{y} \Phi(u) du.
\end{equation}
We now put (\ref{ydot:solution}), which depends on the null
assumption and the equation of motion in the fifth dimension, into
the equations of motion in 4D (\ref{4D:geodesic:1}), to obtain
\begin{equation}\label{4D:geodesic:2}
    {}^{(\lambda)}a^\alpha = - 2K \Phi^{-2} \Phi'k^{\alpha}.
\end{equation}
Also, (\ref{ydot:solution}) back into (\ref{Null:Condition}) gives
\begin{equation}\label{Null:Condition:2}
    K^2 = \Phi^4 k_{\gamma} k^{\gamma}.
\end{equation}
The relations (\ref{4D:geodesic:2}) and (\ref{Null:Condition:2})
describe paths is $(4+1)$D in terms of a parameter
(\ref{y:solution}) which is an integral over the conformal factor
associated with the 4D part of the metric.

Our geodesics depend on three arbitrary parameters $\lambda_0$,
$K$ and $y_0$.  We can remove the former two from the analysis by
performing a transformation of the affine parameter: $\lambda
\rightarrow \tilde{\lambda} = \lambda/|K| + \lambda_0$, provided
$K \ne 0$.  Since $d\tilde{\lambda}/d\lambda > 0$, this
transformation preserves the orientation of the 5D null curve.  We
can include the $K = 0$ case explicitly by defining $\epsilon
\equiv K/|K| = \pm 1$ when $K \ne 0$ and $\epsilon = 0$ when
$K=0$.  Dropping the tilde on the new parameter, we find that
(\ref{y:solution}), (\ref{4D:geodesic:2}) and
(\ref{Null:Condition:2}) become
\begin{eqnarray}\label{First}
    \epsilon\lambda & = & \int_{y_0}^y \Phi(u) du, \\ \label{Second}
    {}^{(\lambda)}a^\alpha & = & -2\epsilon \Phi^{-2} \Phi'k^{\alpha},
    \\ \label{Third}
    \epsilon^2 & = & \Phi^4 k_{\gamma} k^{\gamma}.
\end{eqnarray}
These are the relations we will be concerned with in what follows.
Notice that if $\epsilon = 0$, equation (\ref{First}) implies that
$y = y_0$ for all $\lambda$, which means that there is no motion
in the fifth dimension.

To see where (\ref{First})--(\ref{Third}) lead, let us perform a
parameter transformation $\tau = \tau(\lambda)$.  Then $k^{\alpha}
= v^{\alpha} d\tau/ d\lambda$, where $v^{\alpha} \equiv
dx^{\alpha} / d\tau$. Equations (\ref{Second}) and
(\ref{ydot:solution}) give
\begin{equation}\label{geodesic:transformed}
    {}^{(\tau)}a^\alpha = - \left(
    \frac{d\lambda}{d\tau}
    \right)^{2}  \left(
    \frac{d^2\tau}{d\lambda^2} +
    \frac{2}{\Phi}  \frac{d\Phi}{dy}
    \frac{dy}{d\lambda} \, \frac{d\tau}{d\lambda} \right) v^{\alpha}.
\end{equation}
Clearly, we can choose $\tau = \tau(\lambda)$ in such a way as to
make the right-hand side of (\ref{geodesic:transformed}) zero.
This happens if $d\tau/d\lambda = C/\Phi^2$, where $C$ is a
dimensionless constant we can set equal to unity.  The 4D motion
is described by
\begin{equation}\label{4D:geodesic:3}
    {}^{(\tau)}a^\alpha = 0,
\end{equation}
which is the standard geodesic equation, provided that
\begin{equation}\label{parameter}
    d\tau / d\lambda = \Phi^{-2}.
\end{equation}
This into (\ref{Third}) gives
\begin{equation}\label{proper:time}
    \epsilon^2 = v^{\alpha} v_{\alpha}.
\end{equation}
Hence, we see that $v^{\alpha}(\tau)$ is a 4D geodesic of
$g_{\alpha\beta}$ that can be timelike ($\epsilon = \pm 1$) or
null ($\epsilon = 0$).  Further, this implies that the parameter
$\tau$ is either the proper time or an affine null parameter along
the 4D path in the conformal spacetime described by
$g_{\alpha\beta}$. It is for this reason that we call $\tau$ the
\emph{conformal proper time}. Also, (\ref{ydot:solution}) with
(\ref{parameter}) shows that $dy/d\tau = \epsilon\Phi(y)$, so
\begin{equation}\label{tau:solution}
    \epsilon\tau = \int_{y_0}^y \Phi^{-1}(u)\,du.
\end{equation}
We see that this relation (\ref{tau:solution}) replaces
(\ref{First}), the geodesic (\ref{4D:geodesic:3}) replaces
(\ref{Second}), and (\ref{proper:time}) replaces (\ref{Second}).
In other words, particles which move on null 5D paths have
trajectories that are in accordance with the conventional 4D
geodesic equation, if the parameter is judiciously chosen.  This
brings the physical relevance of the extra force term in equation
(\ref{Second}) into question.  It is important to note from
(\ref{proper:time}) that even though the path is 5D null, it is
not necessarily 4D null. Massive ($v_\alpha v^\alpha
>0$) or massless ($v_\alpha v^\alpha = 0$) particles in 4D can
move on null paths in 5D.

This remarkable result holds irrespective of the form of $\Phi =
\Phi(y)$ in (\ref{restricted:metric}).  However, to make contact
with previous work \cite{Mas94,Mas98,Wes99a,Wes99b}, let us choose
the canonical form $\Phi(y) = y/L$.  Then, $dy/d\tau = \epsilon
y/L$ and $y = y_0 e^{\epsilon\tau/L}$, where $y_0$ is a constant.
By (\ref{restricted:metric}), $d{\mathcal{S}}^2 = 0$ but
$\sqrt{v_\alpha v^\alpha} \ne 0$.

There is yet another choice of parameter that we ought to
consider.  This third parameter choice is based on the induced
metric and is defined by
\begin{equation}\label{hyper}
    \epsilon^2 = h_{\alpha\beta} u^{\alpha} u^{\beta} = \Phi^2
    g_{\alpha\beta} u^{\alpha} u^{\beta},
\end{equation}
where $u^{\alpha} = dx^{\alpha}/ds$.  This parameterization
enforces the proper normalization of the particle trajectory for
observers confined to $\Sigma_y$ hypersurfaces.  It is for this
reason that we call $s$ the \emph{hypersurface proper time}.  The
$s$-parameterization is preferred by Liu and Mashhoon
\cite{Liu00}.  Examining equation (\ref{Third}), we see that we
can satisfy the hypersurface-normalization condition (\ref{hyper})
by setting
\begin{equation}
    ds/d\lambda = \Phi^{-1}.
\end{equation}
Under such a transformation, we can use a formula analogous to
(\ref{geodesic:transformed}) with (\ref{ydot:solution}) to derive
the 4D part of the geodesic equation:
\begin{equation}\label{hyper:geodesic}
    {}^{(s)}a^\alpha = -\epsilon \Phi^{-1} \Phi' u^\alpha.
\end{equation}
We see that in this parameterization we have a velocity-dependent
extra force acting on the particle.  In canonical coordinates
where $\Phi(y) = y/L$, the right-hand side of
(\ref{hyper:geodesic}) becomes $\epsilon u^{\alpha} / (s +
\epsilon)$, i.e.\ it decreases with increasing proper time. This
result represents a deviation from geodesic motion as measured by
observers on $\Sigma_y$. Transforming our solution for $\dot{y}$
(\ref{ydot:solution}) gives
\begin{equation}
    dy/ds = \epsilon \quad \Rightarrow \quad y(s) =
    \epsilon s + y_0.
\end{equation}
That is, the particle has a constant velocity in the $y$ direction
which we have normalized to $\pm 1$  or $0$.  (This is in
agreement with the $K = 0$ case presented by Liu \& Mashhoon
\cite{Liu00}.) We again discover that if the 4D path is null
($\epsilon = 0$), the particle is confined to $\Sigma_y$.

The three types of parameterization that we have discussed in this
section are summarized in Table \ref{tab:summary}.  Of the three
scenarios, the conformal parameterization most resembles what we
are used to in 4D physics.  This might tempt us to decide that the
conformal parameterization is the ``correct'' choice.  However,
such an identification would be premature.  The preferred
parameter in 4D general relativity is the proper time, which has
the geometric interpretation of being the arclength along timelike
geodesics and the physical interpretation of being the time
measured by freely-falling clocks.  In our 5D picture there exists
no useful notion of 5D arclength because the particle trajectories
are null --- we only have the 5D affine parameter $\lambda$. We
have encountered two equally valid notions of 4D arclength: the
proper time in the 4D conformal manifold ($\tau$) and the proper
time associated with the projection of geodesics onto a $\Sigma_y$
hypersurface ($s$).  The only way to distinguish between these
choices is to study the physics associated with each, which is
what we do in the following section.

\section{The physical properties of the trajectories}

When particles follow higher-dimensional geodesic paths, they
often seem to have peculiar physical properties as measured by 4D
observers.  For example, it has been observed by many authors that
particles following geodesic paths in higher dimensions seem to
have variable rest masses according to observers ignorant of the
extra dimensions \cite{Mas98,Liu00}.  We propose to examine the
physical properties of the trajectories derived in the previous
section and hence determine what characteristics of the dynamics
are observationally testable.

\subsection{Rest mass variations and a variable cosmological ``constant''}
\label{sec:Mass}

We want to analyze how an observer ignorant of the fifth dimension
might interpret kinematic data concerning the trajectory of
freely-falling observers in a 5D manifold.  When reducing
observational data, such observers are likely to fall back on the
4D relativistic version of Newton's second law.  That is, they
will demand that the particle's 4-momentum $p^\alpha$ must be
covariantly conserved in a 4D sense:
\begin{equation}\label{2nd:law}
    \frac{Dp^\alpha}{dz} \equiv \frac{dp^\alpha}{dz} +
    \Gamma^{\alpha}_{\beta\gamma} U^\beta p^\gamma =0.
\end{equation}
Here $z$ stands for whatever parameter we are using along the path
($\lambda$, $\tau$, $s$, etc\ldots) and the 4-momentum is assumed
to have the standard form
\begin{equation}
    p^\alpha = mU^\alpha, \quad U^\alpha = \frac{dx^\alpha}{dz},
\end{equation}
where $m$ is the mass.  Let us expand (\ref{2nd:law}), assuming
that the mass varies with $z$.  We obtain
\begin{equation}
    {}^{(z)}a^\alpha = -\frac{1}{m}\frac{dm}{dz}
    U^\alpha.
\end{equation}
If we compare this formula with the results presented in Table
\ref{tab:summary}, we come to a disturbing conclusion: a
particle's mass variation depends explicitly on the choice of
parameterization.  For example, it is easy to see that if the
particle's world line is parameterized by the 5D affine parameter
$\lambda$, the particle mass is given by
\begin{equation}\label{affine:mass}
    m(\lambda) = k \Phi^2\left(y(\lambda)\right),
\end{equation}
where $k$ is a constant.  We can either view the particle mass as
a function of $\lambda$ or as a function of $y$.  If we put
(\ref{affine:mass}) into the normalization condition
(\ref{Third}), we obtain
\begin{equation}
    \epsilon^2 k^{2} = g_{\alpha\beta} (mk^\alpha) (mk^\beta) =
    g_{\alpha\beta} p^\alpha p^\beta.
\end{equation}
Hence, the norm of the four momentum (as defined by the 4D
conformal metric) is conserved along the worldline, which follows
from the fact that $Dp^\alpha / d\lambda = 0$. This is despite the
fact that the norm of the 4-velocity is not constant (the
variation in mass precisely cancels that effect).  Our initial
assumption (\ref{2nd:law}) made no particular choice of 4D metric,
yet $g_{\alpha\beta}$ has been singled out by this calculation.
Now if we chose to raise and lower indices with the induced metric
$h_{\alpha\beta}$, the norm of $p^\alpha$ would be variable,
suggesting that the conformal metric defines the line element
appropriate to observers unaware of the fifth dimension. For the
canonical prefactor $\Phi(y) = y/L$, (\ref{affine:mass}) gives
\begin{equation}
    m(\lambda) = 2kL^{-1}\epsilon^2\lambda,
\end{equation}
where we have chosen $y(\lambda = 0) = 0$.  We see that in the
affine parameterization, the mass increases linearly in ``time''.
However, the variation is small if $L$ is large, or the induced
cosmological constant $\Lambda$ is small.  If $\epsilon = 0$, we
recover that massless particles travel on 4D null geodesics.

Does this interpretation hold up in the hypersurface
parameterization?  The mass function in this case is given by
\begin{equation}
    m(s) = k\Phi(\epsilon s + y_0).
\end{equation}
Again, the mass may be viewed as a function of $y = y(s)$.  The
normalization condition (\ref{hyper}) yields, as before:
\begin{equation}
    \epsilon^2 k^{2} = g_{\alpha\beta} (mk^\alpha) (mk^\beta) =
    g_{\alpha\beta} p^\alpha p^\beta.
\end{equation}
For the canonical prefactor, we obtain
\begin{equation}
    m(s) = kL^{-1} (\epsilon s + y_0).
\end{equation}
The mass is constant if $\epsilon = 0$, and is zero if $k = 0$
also. We note that mass variations are small if the induced
cosmological constant is small.

Finally, we can deal with the trivial case of conformal
parameterizations.  Since the 4D equation of motion
(\ref{4D:geodesic:3}) is precisely affinely geodesic, there is no
mass variation in this parameterization.  This follows from that
fact that the 4-velocity is normalized to have a constant length,
which means that the mass must also be constant to ensure that
$p^\alpha p_\alpha$ is conserved.

We see that there are three different masses for the three
different parameterizations.  However, the conformal metric has
been singled out as the 4D metric appropriate to observers
ignorant of $y$ (as opposed to the induced metric, which is
appropriate to observers confined to $\Sigma_y$ hypersurfaces). In
this parameterization, particle masses are constant.

It therefore becomes obvious that for an arbitrary
parameterization $U^\alpha = dx^\alpha / dz$ related to the affine
parameterization by a transformation of the form $dz/d\lambda =
G(y(\lambda))$, where $G$ is some function of $y$, the mass is
defined by the normalization relation
\begin{equation}
    \epsilon^2 k^2 = m^{2}(y) g_{\alpha\beta} U^\alpha U^\beta,
\end{equation}
where $k$ is a constant.  This has an interesting interpretation
when the 5D vacuum field equations $R_{AB} = 0$ are enforced. From
equation (\ref{Einstein}), the induced 4D stress-energy tensor is
\begin{equation}
    8\pi T_{\alpha\beta} = \Lambda g_{\alpha\beta}.
\end{equation}
Now, the energy density $\rho$ of cosmological matter will be
measured by an observer with 4-velocity $U^\alpha$ to be
\begin{equation}
    8\pi\rho = 8\pi T_{\alpha\beta} U^\alpha U^\beta
    = \epsilon^2 \Lambda k^2 m^{-2}(y).
\end{equation}
Hence, there is a direct relation between the energy density of
the vacuum and the mass of the particle.  If the particle mass
varies, an observer traveling along with the particle will measure
the energy density of the vacuum to be variable.  That is, the
observer will measure a variable cosmological ``constant''.  We
can consider small changes in the particle mass $\delta m$
connected with small changes in the energy density $\delta \rho$:
\begin{equation}
    \delta m = -4\pi \Lambda^{-1} k^{-2} m^3 \delta \rho,
\end{equation}
where we have taken $\epsilon^2 = 1$.  This has the suggestive
form of an energy conservation equation. Let us assume that the
particle has a 3D ``volume'' associated with it that is related to
its mass $V = V(m)$ [as in the black hole case]. Let us also
assume that a change in vacuum energy $\delta E$ in the volume
occupied by the particle results in an increase or decrease of the
particle mass: $\delta m = -\delta E$.  However, we have $\delta E
= \delta(\rho V)$. Using these relations we can derive a
differential equation for $dV/dm$:
\begin{equation}
    0 = m \, dV/dm - 2V - 8\pi\Lambda^{-1}k^{-2} m^3,
\end{equation}
where we have cancelled a common factor of $\delta m$.  Setting
$V(m=0)=0$, which implies that massless particles remain massless,
we get
\begin{equation}
    V(m) = 8\pi\Lambda^{-1}k^{-2} m^3.
\end{equation}
Therefore, as particles move through the 5D manifold they will in
general observe the cosmological ``constant'' to be varying in
time. Further, if one assumes that the particle occupies a 3D
volume of linear dimension $\sim m$ (as is the case for a black
hole) then the energy being gained or lost by the vacuum
corresponds to the decrease or increase of the particle's mass.

\subsection{Constants of the motion and the particle
energy}\label{sec:Conserved}

The fact that the definition of rest mass is parameter-dependent
may be considered by some to be unsatisfactory.  A physical
quantity like $m$ should be independent of the timing mechanism
employed to separate points along the particle's worldline.  To
remove the ambiguity in parameterization, we attempt to construct
observable quantities that depend only on the 5D coordinates and
not the parameter.  A physically meaningful class of observables
for spacetimes with a certain degree of symmetry are the constants
of the motion, such as the energy, linear momentum, angular
momentum, etc\ldots.  We can argue that such quantities ought to
be independent of $y$, which implies that there is no intrinsic
rest-mass variation.

Let us assume that the conformal manifold admits the existence of
a Killing vector $\eta^\alpha$ such that $\pounds_\eta
g_{\alpha\beta} = 0$.  Then by the results of section
\ref{sec:Killing}, the 5D manifold has a Killing vector of the
form
\begin{equation}\label{up}
    \xi^A = (\eta^\alpha,0).
\end{equation}
We exclude the special case $\Omega(y) = \Omega_0 \exp(k_3y)$, so
all 5D Killing vectors are of the form (\ref{up}).  Now, since
$k^B \nabla_B k^A = 0$, we have that
\begin{equation}
    {\mathcal{K}}_\eta = \xi_A k^A
\end{equation}
is a constant of the motion.  Here $k^A = dx^A / d\lambda$.  We
would like to write ${\mathcal{K}}_\eta$ in a form independent of
the parameter.  To do so, we introduce a time foliation of the 4D
part of the manifold.  This allows us to write the conformal line
element in lapse and shift form:
\begin{eqnarray}\nonumber
    d\tau^2 & = & g_{\alpha\beta}dx^\alpha dx^\beta \\
    & = & g_{00} \left[ \, dt^2 -
    \sigma_{ij} (N^{i}\,dt + dx^{i}) (N^{j}\,dt + dx^{j}) \right],
\end{eqnarray}
where $i,j=1,2,3$.  Here, $g_{00}$ is the redshift factor, $N^{i}$
is the shift 3-vector and $g_{00}\sigma_{ij}$ is the 3-metric.  We
can use the normalization condition for the affine
parameterization (\ref{Third}) (with $\epsilon = 1$) to obtain
\begin{equation}
    \frac{d\lambda}{dt} = \Phi^{2}(y) g^{1/2}_{00} (1-\beta^2)^{1/2},
\end{equation}
where
\begin{equation}
    \beta^2 = \sigma_{ij} (N^{i} + V^{i}) (N^{j} + V^{j}),
\end{equation}
with $V^i = dx^i /dt$.  By an appropriate choice of foliation we
can set $N^i = 0$, which reduces $(1-\beta^2)^{1/2}$ to the
Lorentz factor $(1-V^2)^{1/2}$ when $g_{00} = 1$.  Therefore, we
may write ${\mathcal{K}}_\eta$ as
\begin{equation}\label{constant}
    {\mathcal{K}}_\eta = \frac{g_{\alpha\beta} \eta^\alpha
    V^\beta}{g^{1/2}_{00} (1-\beta^2)^{1/2}}, \quad V^\alpha =
    \frac{dx^\alpha}{dt}.
\end{equation}
This form is independent of the parameter choice used to solve the
5D geodesic equation.  It is also independent of the extra
dimension $y$ by virtue of the fact that $g_{\alpha\beta}$ and
$\eta^\alpha$ are functions of spacetime variables only.
Therefore, if observers can measure the 4-dimensional position of
a freely-falling particle at different points along its world
line, they can construct the constants of the motion without
knowledge of $dy/d\lambda$ or $dy/dt$.

We can illustrate this point by considering a specific 4D metric
which corresponds to a solution of the 5D vacuum field equations,
namely the Schwarzschild--de Sitter one:
\begin{eqnarray}\nonumber
    g_{\alpha\beta} dx^\alpha dx^\beta & = & f(r)\,dt^2 - f^{-1}(r) \,
    dr^2 - r^2 \, d\Omega^2, \\
    \label{de Sitter} f(r) & = & 1 - \frac{2M}{r} - \frac{\Lambda}{3}
    r^2.
\end{eqnarray}
Here, $M$ is the mass while $\Lambda$ is the induced cosmological
constant (\emph{cf}.\ section \ref{sec:Unique}).  This spacetime
is static and spherically symmetric, so it has a timelike Killing
vector $\tau^\alpha = (\di / \di t)^\alpha$ that we can use to
define the energy, and an azimuthal Killing vector $\phi^\alpha =
(\di / \di \phi)^\alpha$ we can use to define the angular
momentum.  For an equatorial orbit ($\theta = \pi/2$), the angular
momentum (up to a multiplicative constant) is
\begin{equation}\label{L}
    L = \frac{f^{-1/2}(r) r^2}{(1-\beta^2)^{1/2}}
    \frac{d\phi}{dt},
\end{equation}
where $\beta^2 = r^2 f^{-1}(r) d\phi / dt$.  If the mass of the
particle varies with $y$, we would expect an additional function
of $y$ to appear in this expression.  That is, if $m$ changes as
the particle moves between $\Sigma_y$ hypersurfaces, then we
physically expect either the orbital velocity $d\phi/dt$ or the
particle's radial position $r$ to change in a fashion that leaves
$L$ constant. However, equation (\ref{constant}) implies that the
particle's 4D worldline is insensitive to motion in the $y$
direction, which is a direct consequence of the conformal equation
of motion (\ref{4D:geodesic:3}). This tends to support the view
that the particle mass is constant. This argument is not
restricted to spacetimes with azimuthal symmetry, since the
general form of the constants of motion (\ref{constant}) is
general.

Using $\tau^\alpha = (\di / \di t)^\alpha$, we can also define the
particle's energy. Let us take $1 \gg 2M/r \gg \Lambda r^2 / 3$
and consider only radial motion $d\theta / dt = d\phi /dt = 0$.
Then to first order in $M$ and $v_r = dr/dt$, and zeroth order in
$\Lambda$, the energy is
\begin{equation}
    E = 1 + \frac{v_r^2}{2} - \frac{M}{r} + \cdots
\end{equation}
The second and third terms are obviously the Newtonian kinetic and
potential energies, which means that the first term must be the
rest mass energy.  The fact that the rest mass energy is a
constant independent of $y$ confirms that the rest mass does not
vary along the particle's world line, at least according to an
analysis based on constants of the motion.

\subsection{5D time dilation}\label{sec:Time}

While the calculation of the previous subsection has the advantage
of being independent of the parameterization of the trajectory, it
has the disadvantage of being a coordinate-dependent manipulation
that relies heavily on our choice of foliation.  That is, the $dt$
coordinate time interval is not an invariant quantity. However,
the 4D proper time interval is indeed an invariant under 4D
coordinate transformations, which suggests that we ought to write
our equations in terms of $dz$.  The problem is that each of the
parameterizations $\lambda$, $\tau$ or $s$ (and others) could
qualify as the proper time $z$. To our knowledge, there is no
\emph{a priori} method for determining the ``true'' 4D proper
time; but it is easy to imagine an experiment which would show
which one is most convenient.

Consider a spherically-symmetric spacetime that allows for
circular orbits [like (\ref{de Sitter}) above].  By virtue of
equations (\ref{4D:geodesic:3}) and (\ref{proper:time}), the
conformal time interval $\Delta\tau$ associated a complete
revolution in a circular orbit is constant (i.e.\ the orbital
velocity $d\phi / d\tau$ is constant).  Now, suppose that we have
a satellite in a circular orbit that carries an atomic clock or
some other time-keeping device.  This clock measures the proper
time along the path by ``ticking'' $\Delta N / \gamma$ times
during a proper time interval $\Delta z$ ($\gamma$ is the constant
rate at which the clock oscillates).  As seen above, the relation
between different 4D parameterizations is in general given by
\begin{equation}
    d\tau / dz = F(y(\tau)),
\end{equation}
where $F$ is some function of $y$, so that
\begin{equation}
    \Delta N = \gamma \int_{\tau_i}^{\tau_i + \Delta \tau} F^{-1}(y(\tau))
    \, d\tau.
\end{equation}
Here $\tau_i$ is when we start keeping time and also represents
the initial $y$ position of the circular orbit.  Now, let us count
the number of oscillations $\Delta N_1$ that our clock undergoes
during a complete orbit starting at time $\tau_1$, and then repeat
the procedure for another orbit starting at a later time $\tau_2$.
If we adopt the canonical prefactor $\Phi(y) = y/L$, the ratio of
the number of clock oscillations during the two orbits is
\begin{equation}
    \frac{\Delta N_2}{\Delta N_1} = \left\{
    \begin{array}{ll}
        e^{2\epsilon (\tau_2 - \tau_1) / L}, & z=\lambda \\
        1, & z=\tau \\
        e^{\epsilon (\tau_2 - \tau_1) / L}, & z=s.
    \end{array}
    \right.
\end{equation}
Here the time parameter $\tau$ is related to $y$ via
\begin{equation}
    y(\tau) = y_0 e^{\epsilon\tau/L}.
\end{equation}
Therefore, if the 4D proper time is not $\tau$ then the time
elapsed in the rest frame of the orbiting body during one complete
revolution will not be constant.  That is, an observer moving with
the clock will conclude that the clock is speeding up or slowing
down (depending on whether they are moving in the direction of
increasing or decreasing $y$); or that the orbital velocity $d\phi
/ dz \approx 2\pi\gamma / \Delta N$ is growing smaller or larger
with time.  Of course, the effect is small if $L$ is large or
$\Lambda$ is small.  These effects are in principle testable, and
could be used to distinguish between possible candidates for the
best proper time.

\section{Conclusion}

To better understand the dynamics of particles moving in a
higher-dimensional world but observed in spacetime, we have
introduced a 5D warped-product space which is related to the 4D
sub-manifold via a conformal factor that depends on the extra
coordinate. When the 5D field equations are the standard (vacuum)
ones of Kaluza-Klein theory, the 4D sub-manifold represents
spacetimes with a non-zero cosmological constant and devoid of
ordinary matter.  We have examined the 5D Killing vectors, which
are related to the 4D Killing vectors of the sub-manifold. A major
result is that null geodesics in 5D can correspond to non-null
geodesics in 4D. That is, massless particles in Kaluza-Klein space
can correspond to massive particles in Einstein space.  It has
been shown that there is in general an anomalous acceleration in
the 4D equation of motion that can be removed by a parameter
transformation.  This brings up the question of how the ``true''
proper time, which is the time measured by freely falling clocks,
should be chosen. The ambiguity in the choice of parameterization
results in multiple expressions for the particle mass, which in
general vary along the particle's worldline.  In the 5D vacuum
case, the variation in rest mass can be related to the variation
in the vacuum energy as measured by an observer traveling with the
particle.  However, we have shown that the constants of the motion
can be written in a form independent of both the parameter and the
extra coordinate, which suggests to us that variable rest mass may
be an artifact of a poor choice of parameter.  We have argued that
the best choice of the parameter that describes a particle's
motion, and the question of the variability of its rest mass, can
be tested by experiment or observation.

In closing, we should remind ourselves that modern Kaluza-Klein
theory (without the cylinder and compactification condition) is
fully covariant in 5D.  One can argue that the same requirements
be made of superstrings in 10D and supergravity in 11D, and this
is indeed a strength of much recent work on brane theory in $N$D.
However, we currently interpret experimental and observational
data in terms of four spacetime dimensions.  Therefore, to make
contact with everyday experience we naturally attempt to
interpret 5D geometric objects, like null geodesics, within the
context of 4D spacetime.  This dimensional reduction is the
reason that massless particles in 5D can appear to have (possibly
variable) finite rest masses in 4D.  However, the details of the
reduction from 5D to 4D are not unique, so we suggest that
further work be done to determine the most convenient reduction
scheme.

\begin{acknowledgements}
We would like to thank W. N. Sajko for useful comments, and NSERC
of Canada for financial support.
\end{acknowledgements}

\newpage

\mediumtext
\begin{table}

\caption{Summary of the different type of parameterization of 5D
null geodesics}

\label{tab:summary}

\begin{tabular}{llll}

Parameterization & 4D Normalization & 4D Equation of Motion &
Motion in $y$ direction
\\ \tableline

Affine ($\lambda$) & $\epsilon^2 = \Phi^4 g_{\alpha\beta}
k^{\alpha} k^{\beta}$ & ${}^{(\lambda)}a^\alpha =
-2\epsilon\Phi^{-2} \Phi' k^{\alpha}$ & $\epsilon\lambda = \int^y
\Phi(u) \, du$

\\

Conformal ($\tau$) & $\epsilon^2 = g_{\alpha\beta} v^{\alpha}
v^{\beta}$ & ${}^{(\tau)} a^\alpha = 0$ & $\epsilon\tau = \int^y
\Phi^{-1}(u)\,du$

\\

Hypersurface ($s$) & $\epsilon^2 = h_{\alpha\beta} u^{\alpha}
u^{\beta}$ & ${}^{(s)}a^\alpha = -\epsilon \Phi^{-1} \Phi'
u^\alpha$ & $y(s) = \epsilon s + y_0$

\end{tabular}

\end{table}


\begin{references}

\bibitem{Ran99}
    L. Randall, R.Sundrum, {\tt{hep-th/9906064}} (1999).

\bibitem{Shi99}
    T. Shiromizu, K. Maeda, M. Sasaki, {\tt{gr-qc/9910076}} (1999).

\bibitem{Dzh99}
    V. Dzhunushaliev, H. -J. Schmidt, Grav. Cosmol., 5, 187
    (1999).

\bibitem{Wes00}
    P. S. Wesson, H. Liu, S. S. Seahra, Astronomy \& Astrophysics,
    358, 425 (2000).

\bibitem{Lei73}
    E. Leibowitz, N. Rosen, General Relativity and Gravity, 4, 449
    (1973).

\bibitem{Kov84}
    D. Kovacs, General Relativity and Gravity, 16, 645 (1984).

\bibitem{Geg84}
    J. Gegenberg, G. Kunstatter, Physics Letters A, 106, 410
    (1984).

\bibitem{Dav86}
    A. Davidson, D. A. Owen, Physics Letters B, 177, 77 (1986).

\bibitem{Fer89}
    J. A. Ferrari, General Relativity and Gravity, 21, 683 (1989).

\bibitem{Mas94}
    B. Mashhoon, H. Liu, P. S. Wesson, Physics Letters B, 331, 305
    (1994).

\bibitem{Wes95}
    P. S. Wesson, J. Ponce de Leon, Astronomy \& Astrophysics,
    294, 1 (1995).

\bibitem{Mas98}
    B. Mashhoon, P. S. Wesson, H. Liu, General Relativity and
    Gravitation, 30, 555 (1998).

\bibitem{Wes99a}
    P. S. Wesson, B. Mashhoon, H. Liu, W. N. Sajko, Physics
    Letters B, 456, 34 (1999).

\bibitem{Wes99b}
    P. S. Wesson, Space-Time-Matter, World Scientific, Singapore
    (1999).

\bibitem{Liu00}
    H. Liu, B. Mashhoon, {\tt{gr-qc/0005079}} (2000).

\bibitem{Wil92}
    C. M. Will, International Journal of Modern Physics D, 1, 13
    (1992).

\bibitem{Kal21}
    T.\ Kaluza, Stiz.\ Preuss.\ Akad.\ Wiss.\ Math.\ Kl., 966 (1921).

\bibitem{Eis49}
    L. P. Eisenhart, Riemannian Geometry, Princeton University
    Press, Princeton (1949).

\bibitem{Syn49}
    J. L. Synge, A. Schild, Tensor Calculus, Dover Publications,
    New York (1949).

\bibitem{Wal84}
    R. M. Wald, General Relativity, University of Chicago Press,
    Chicago (1984).

\end{references}
\end{document}